\documentclass[pra,twocolumn,showpacs,preprintnumbers,amsmath,amssymb]{revtex4}
\usepackage{mathrsfs}
\usepackage{bbm}
\usepackage{amsfonts}
\usepackage{tipa}

\usepackage{epsfig,graphicx}
\usepackage{amstext}
\usepackage{amsmath}
\usepackage{graphicx}

\begin{document}
%%%%%%%%%%%%%%%%%%%%%%%%%%%%%%%%%%%%%%%%%%%%%%%%%%%%%%%%%%%%%%%%%%%%%%

%TCIDATA{OutputFilter=Latex.dll}
%TCIDATA{Version=5.00.0.2552}
%TCIDATA{<META NAME="SaveForMode" CONTENT="1">}
%TCIDATA{LastRevised=Wednesday, June 22, 2005 16:21:09}
%TCIDATA{<META NAME="GraphicsSave" CONTENT="32">}

\title{Improving teleportation fidelity in structured reservoirs}

\author{Yu-Xia Xie}
\email{yuxia1124@163.com}
\author{Xiao-Qiang Xi}
\address{School of Science, Xi'an University of Posts and Telecommunications, Xi'an 710121, China}

\begin{abstract}
Seeking flexible methods to control quantum teleportation in open
systems is an important task of quantum communication. In this
paper, we study how the super-Ohmic, Ohmic and sub-Ohmic reservoirs
affect teleportation of a general one-qubit state. The results
revealed that the structures of the reservoirs play a decisive role
on quality of teleportation. Particularly, the fidelity of
teleportation may be improved by the strong backaction of the
non-Markovian memory effects of the reservoir. The physical
mechanism responsible for this improvement are determined.
\end{abstract}


\maketitle

\section{Introduction}
Since the seminal work of Bennett et al. \cite{Bennett}, quantum
teleportation undergoes rapid development both in the theory
\cite{Lee,Bowen,Verstraete,Yeoprl,Hupra,Huml} and experiments
\cite{Bouwmeester,Nielsen,Furusawa,Olmschenk}. Particularly, quantum
teleportation via free-space channels over 97 kilometres
\cite{Panjw} and 143 kilometres \cite{Zeilinger} have been
experimentally realized recently by two research groups leaded by
Pan and Zeilinger, respectively. These finding rekindled the dream
of people on constructing ultra-long-distance quantum communication
networks on a global scale. To make this dream come true, one needs
to face many actual challenges. One such challenge is the
decoherence of quantum states due to its inevitable interaction with
the surroundings \cite{Almeida,Xujs,Liubh}. This unwanted
interaction in most cases causes degradation of quantum correlations
\cite{Mazzola,Huaop1,Huaop2}, and therefore limits the fidelity as
well as the distance of quantum teleportation. Even for the simplest
protocol of one-qubit teleportation \cite{Bennett}, there are many
processes that decoherence may be set in, e.g., the preparation and
distribution processes of the shared quantum channels, the imperfect
measurements of the information sender (Alice) and receiver (Bob).
All these make it significant and vital to investigate resistance of
a teleportation protocol under the influence of decoherence, and
this subject has been studied by many authors from the perspective
of both noise channels
\cite{Oh,Jung,Rao,Hupla1,Hupla2,Hujpb1,Hujpb2} and noisy operations
\cite{Yeoepl,Huepjd}. Particularly, it has been shown that although
entanglement is the necessary resource for quantum teleportation, in
general its magnitude is not proportional to the value of the
teleportation fidelity \cite{Hupla2,Hujpb1,Hujpb2}.

In the present work, we study quantum teleportation of the
single-qubit state in the bosonic structured reservoirs of the
super-Ohmic, Ohmic and sub-Ohmic types \cite{Leggett}. Through
simulating numerically the dynamical behaviors of the average
teleportation fidelity, we show that the explicit form of the
spectral density of the reservoir as well as the coupling strength
between the channel and the reservoir have great impact on
efficiency of quantum teleportation, and the quality of
teleportation can be improved via the technique of reservoir
engineering.

The structure of this paper is arranged as follows. In Section II we
recall some basic formalism related to the fidelity of quantum
teleportation, and detailed the methods for solving the equation of
motion for the channel state. Then in Section III, we investigate
resistance of quantum teleportation against decoherence induced by
the super-Ohmic, Ohmic and sub-Ohmic reservoirs. Finally, Section IV
is devoted to a summary.

\section{The formalism}
In this work we concentrate on efficiency of quantum teleportation
for a general one-qubit state, with a two-qubit state $\rho$ serving
as the quantum channel. We suppose during the teleportation process,
Alice performs only the Bell-basis measurement, while Bob is
equipped to perform any unitary transformation, then the maximal
average fidelity achievable can be evaluated as \cite{Horodeckiletta}
\begin{equation}\label{eq1}
 F_{\rm av}=\frac{1}{2}+\frac{1}{6}N(\rho),
\end{equation}
where $N(\rho)={\rm Tr}\sqrt{T^\dag T}$, with $T$ being the $3\times
3$ positive matrix with elements $t_{nm}$ related to the Bloch
sphere representation of $\rho$ below
\begin{equation}\label{eq2}
 \rho=\frac{1}{4}\sum_{n,m=0}^3 t_{nm}\sigma_n\otimes\sigma_m,
\end{equation}
with $\sigma_0$ being the $2\times2$ identity operator,
$\sigma_{1,2,3}$ are the usual Pauli matrices, and $t_{nm}={\rm
Tr}(\rho\sigma_n\otimes\sigma_m)$.

To evaluate resistance of the teleportation protocol, we suppose the
channel state consists of two identical qubits which coupled
independently to their own reservoir, with the single
``qubit+reservoir'' Hamiltonian reads \cite{Breuer}
\begin{equation}\label{eq3}
 {\hat H}=\omega_0\sigma_{+}\sigma_{-}
          +\sum_{k}\omega_k b_k^{\dag} b_k+\sum_{k}(g_k b_k\sigma_{+}+{\rm H.c.}),
\end{equation}
with $\omega_0$ being the Bohr frequency of the two qubits, and the
index $k$ labels the reservoir field mode with frequency $\omega_k$
and the system-reservoir coupling strength $g_k$. Moreover,
$\sigma_+$ ($\sigma_{-}$) are the usual Pauli raising (lowering)
operator, while $b_k^{\dag}$ ($b_k$) are the bosonic creation
(annihilation) operator.

Under the assumption of zero-temperature reservoir and nullity of
initial correlation between the qubit and the reservoir, the reduced
density matrix $\rho^S(t)$ for qubit $S$ ($S=A,B$) can be obtained
as \cite{Breuer}
\begin{equation}\label{eq4}
 \rho^S(t)=\left(\begin{array}{cc}
           \rho^S_{11}(0)|p(t)|^2  & \rho^S_{10}(0)p(t) \\
           \\
           \rho^S_{01}(0)p^*(t) & 1-\rho^S_{11}(0)|p(t)|^2
    \end{array}\right),
\end{equation}
in the standard basis $\{|1\rangle,|0\rangle\}$, and
$\rho^S_{ij}(0)=\langle i|\rho^S(0)|j\rangle$. The density matrix
$\rho^S(t)$ depends solely on the function $p(t)$, whose explicit
time dependence contains the information on the reservoir spectral
density and the coupling constants. To be explicitly, we consider
here the family of reservoir spectral densities given by
\cite{Leggett}
\begin{equation}\label{eq5}
 J(\omega)=\sum_{k}|g_k|^{2}\delta (\omega-\omega_k),
\end{equation}
which simplifies to
\begin{equation}\label{eq6}
 J(\omega)=\eta\omega^s\omega_c^{1-s} e^{-\omega/\omega_c},
\end{equation}
in the continuum limit of reservoir modes
$\sum_{k}|g_k|^{2}\rightarrow\int d(\omega)J(\omega)$, with
$\omega_c$ being the cutoff frequency and $\eta$ the dimensionless
coupling constant. They are related to the reservoir correlation
time $\tau_B$ and the relaxation time $\tau_R$ by
$\tau_B\simeq\omega_c^{-1}$ and $\tau_R\simeq\eta^{-1}$. By tuning
the value of $s$, the spectral densities may be in the regime of
sub-Ohmic $(0<s<1)$, Ohmic $(s=1)$ or super-Ohmic $(s>1)$.

For this family of reservoir spectral densities, the function $p(t)$
obeys the following integro-differential equation \cite{Tongqj}
\begin{equation}\label{eq7}
 \dot{p}(t)+i\omega_{0}p(t)+\int_0^t p(t_1)f(t-t_1)dt_1=0,
\end{equation}
where the kernel function $f(t-t_1)=\int d\omega
J(\omega)e^{-i\omega(t-t_1)}$. After a straightforward algebra, Eq.
\eqref{eq7} turns into
\begin{equation}\label{eq8}
  p(t)=p(0)-\int_{0}^{t}\left(i\omega_0+\int_{t_1}^{t}f(t-t_1)dt\right)p(t_1)dt_1.
\end{equation}
In this work we take $s=1/2$, 1 and 3 as three examples of the
sub-Ohmic, Ohmic, and super-Ohmic spectral densities. The kernel
function can then be integrated as \cite{Hupra}
\begin{equation}\label{eq9}
  f(t-t_1)=\left\{
    \begin{aligned}
         &\frac{s!\eta\omega_c^2}{[1+i\omega_c(t-t_1)]^{s+1}}\quad (s\in\mathbb{Z}),\\
         &\frac{\sqrt{\pi}\eta\omega_c^2e^{-i\varpi}}{2[1+\omega_c^2 (t-t_1)^2]^{3/4}}\quad(s=1/2),
    \end{aligned} \right.
\end{equation}
where $s!$ is the factorial of $s$, and
$\varpi=\frac{3}{2}\tan^{-1}[\omega_c (t-t_1)]$. With the help of
Eq. \eqref{eq9}, the function $p(t)$ in Eq. \eqref{eq8} can be
solved numerically and the two-qubit density matrix $\rho(t)$ can be
determined by the procedure of Ref. \cite{Bellomoprl}.

In fact, the equation of motion of qubit $S$ can also be written in
the following master equation form (by differentiating Eq.
\eqref{eq4} with respect to time) of the form
\begin{eqnarray}\label{eq10}
 \dot{\rho}^{S}(t)&=&-i\frac{\Omega(t)}{2}[\sigma_{+}\sigma_{-},\rho^{S}(t)]+
                     \frac{\Gamma(t)}{2}[2\sigma_{-}\rho^{S}(t)\sigma_{+}\nonumber\\
                   &&-\sigma_{+}\sigma_{-}\rho^{S}(t)-\rho^{S}(t)\sigma_{+}\sigma_{-}],
\end{eqnarray}
with the time-dependent decay rate $\Gamma(t)$ and Lamb shift
$\Omega(t)$ given by \cite{Breuer}
\begin{eqnarray}\label{eq11}
 \Gamma(t)=-2{\rm Re}\left[\frac{\dot{p}(t)}{p(t)}\right],~
 \Omega(t)=-2{\rm Im}\left[\frac{\dot{p}(t)}{p(t)}\right].
\end{eqnarray}
In general, the absolute values of $\Gamma(t)$ may be larger than
unity. In the following, we define the normalized decay rate
$\gamma(t)=\Gamma(t)/\Gamma_{\rm\max}(t)$ for better visualizing the
corresponding phenomena, where the maximum is taken over the entire
time region for fixed $\eta$ and $\omega_c$.

\section{Average fidelity dynamics}

Here we suppose the channel state is prepared initially in the
Werner-like state \cite{Werner}
\begin{eqnarray}\label{eq12}
 \rho(0)=r|\Psi\rangle\langle\Psi|+\frac{1-r}{4}\mathbb{I},
\end{eqnarray}
where $|\Psi\rangle=(|00\rangle+|11\rangle)/\sqrt{2}$. $\rho(0)$ is
entangled if $1/3<r \leqslant 1$ and separable if $0\leqslant
r\leqslant 1/3$. For this kind of $\rho(0)$, we have
\begin{equation}\label{eq13}
 N(\rho)=(r+1)|p(t)|^4+2(r-1)|p(t)|^2+1,
\end{equation}
therefore the average teleportation fidelity $F_{\rm av}$ is
determined by the absolute values of $p(t)$, and this applies to all
kinds of reservoir spectral densities with the Hamiltonian model of
Eq. \eqref{eq3}.

\begin{figure}
\centering
\resizebox{0.35\textwidth}{!}{%
\includegraphics{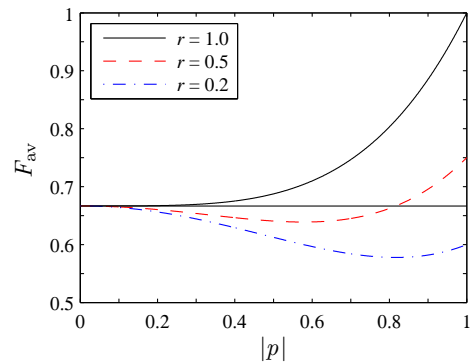}}
% If not, use\vspace{5cm}
% Give the correct figure height in cm
\caption{(Color online) The $|p(t)|$ dependence of the average
         fidelity $F_{\rm av}$ with different mixing parameters of
         $\rho(0)$.} \label{fig:1}
% Give a unique label
\end{figure}

By combining Eqs. \eqref{eq1} and \eqref{eq13}, one can identify
three different types of behavior of the $|p|$ dependence of $F_{\rm
av}$ (see, Fig. \ref{fig:1}), which depend on the values of the
mixing parameter $r$: (i) for $r=1$, $F_{\rm av}$ decays
monotonically with the decrease of $|p|$ and arrives at the
classical limiting value of $2/3$ when $|p|=0$; (ii) for $1/3<r<1$,
$F_{\rm av}$ first decays with the decrease of $|p|$, and then
increase with the decrease of $|p|$ after the critical point
$|p|_{c}^2=(1-r)/(1+r)$, and $F_{\rm av}$ is larger than $2/3$ only
when $|p|^2>2|p|_{c}^2$; (iii) for $0\leqslant r\leqslant 1/3$,
although $F_{\rm av}$ still decays with the decrease of $|p|$, and
increase with the decrease of $|p|$ after $|p|>|p|_{c}$, its value
cannot exceed $2/3$.

\begin{figure}
\centering
\resizebox{0.35\textwidth}{!}{%
\includegraphics{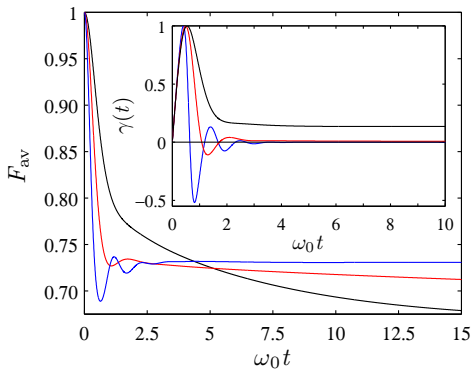}}
% If not, use\vspace{5cm}
% Give the correct figure height in cm
\caption{(Color online) Average fidelity $F_{\rm av}$ versus
         $\omega_{0}t$ in the super-Ohmic reservoir with $\omega_c=\omega_0$, $r=1$, and
         $\eta=0.15$ (black), 0.35 (red), 0.9 (blue). The inset shows behaviors of
         the corresponding normalized decay rates $\gamma (t)$.} \label{fig:2}
% Give a unique label
\end{figure}

In Fig. \ref{fig:2} we display dynamical behaviors of the average
teleportation fidelity $F_{\rm av}$ for the case of super-Ohmic
reservoir with the cutoff frequency $\omega_c=\omega_0$, mixing
parameter $r=1$ and different coupling strengths $\eta$. One can see
that in the weak-coupling regime (e.g., $\eta=0.15$), $F_{\rm av}$
decays monotonically and approaches to its limiting value 2/3 only
when $\omega_{0}t\rightarrow \infty$. If one enlarges the coupling
strength moderately (e.g., $\eta=0.35$), the strong non-Markovian
effect dominates and this leads to an obvious oscillation of $F_{\rm
av}$, but in the long-time limit it still reaches the classical
limiting value 2/3. If one further increases the coupling strength
to a much larger value (e.g., $\eta=0.9$), however, $F_{\rm av}$
firstly experiences some oscillations, and then approaches to a
finite value larger than 2/3 in the long-time limit.

To understand the physical mechanism responsible for the above
phenomena, we plot in the inset of Fig. \ref{fig:2} the behaviors of
the normalized decay rate $\gamma(t)$. Clearly, for $\eta=0.15$ the
decay rate maintains a finite value after one oscillation, and thus
deprives the quantum advantage of the teleportation protocol in the
long-time limit. For $\eta=0.9$, however, $\gamma(t)$ approaches
zero after some oscillations due to the energy and/or information
exchanging back and forth between the qubits and their independent
memory reservoirs, and this leads to $F_{\rm av}>2/3$ even in the
long-time limit. The origin of $F_{\rm av}(t\rightarrow\infty)>2/3$
for $\eta=0.9$ can also be explained from the formation of the bound
state in the system. In fact, for the Hamiltonian model of Eq.
\eqref{eq3} with the spectral densities of the form of Eq.
\eqref{eq6}, the conditions for formation of a bound state can be
determined as
\begin{equation}\label{eq14}
  \eta>\left\{
    \begin{aligned}
         &\frac{\omega_0}{(s-1)!\omega_{c}}\quad (s\in\mathbb{Z}),\\
         &\frac{\omega_0}{\sqrt{\pi}\omega_{c}}\quad(s=1/2),
    \end{aligned} \right.
\end{equation}
by the same methodology of Ref. \cite{Tongqj}. For the system
parameters chosen in Fig. \ref{fig:2} (i.e., $s=3$ and
$\omega_c=\omega_0$), $\eta=0.9$ is clearly larger than its critical
value 0.5, and therefore $F_{\rm av}(t\rightarrow\infty)>2/3$. For
$\eta=0.15$ and 0.35, which are smaller than 0.5 and thus $F_{\rm
av}(t\rightarrow\infty)=2/3$, although there may be weak enhancement
of $F_{\rm av}$ (e.g., $\eta=0.35$) during certain $\omega_0 t$
regions caused by the non-Markovian effects.

\begin{figure}
\centering
\resizebox{0.35\textwidth}{!}{%
\includegraphics{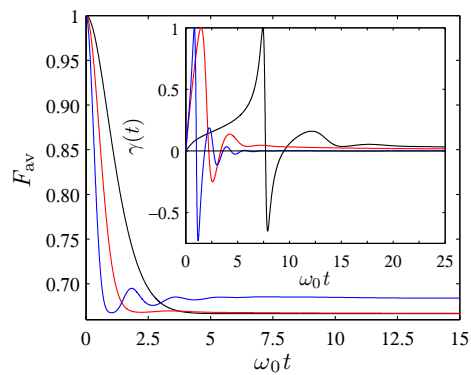}}
% If not, use\vspace{5cm}
% Give the correct figure height in cm
\caption{(Color online) Average fidelity $F_{\rm av}$ versus
         $\omega_{0}t$ in the Ohmic reservoir with $\omega_c=\omega_0$, $r=1$, and
         $\eta=0.3$ (black), 0.9 (red), 2.7 (blue). The inset shows behaviors of the
         corresponding normalized decay rates $\gamma (t)$.} \label{fig:3}
% Give a unique label
\end{figure}

\begin{figure}
\centering
\resizebox{0.35\textwidth}{!}{%
\includegraphics{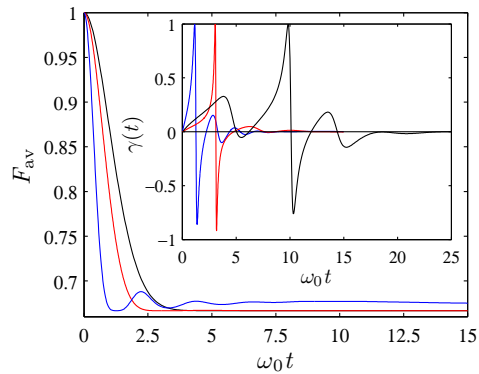}}
% If not, use\vspace{5cm}
% Give the correct figure height in cm
\caption{(Color online) Average fidelity $F_{\rm av}$ versus
         $\omega_{0}t$ in the sub-Ohmic reservoir with $\omega_c=\omega_0$, $r=1$, and
         $\eta=0.3$ (black), 0.55 (red), 2.1 (blue). The inset shows behaviors of
         the corresponding normalized decay rates $\gamma (t)$.} \label{fig:4}
% Give a unique label
\end{figure}

When considering the Ohmic (Fig. \ref{fig:3}) and sub-Ohmic (Fig.
\ref{fig:4}) reservoirs, similar phenomena are observed (the
oscillating behaviors of $F_{\rm av}$ for the Ohmic reservoir with
$\eta=0.9$ and sub-Ohmic reservoir with $\eta=0.55$ are very weak
and unobvious in the plots), with however different critical
coupling strengths for the formation of a bound state. This
corroborates the observation derived for the case of super-Ohmic
reservoir.

\begin{figure}
\centering
\resizebox{0.35\textwidth}{!}{%
\includegraphics{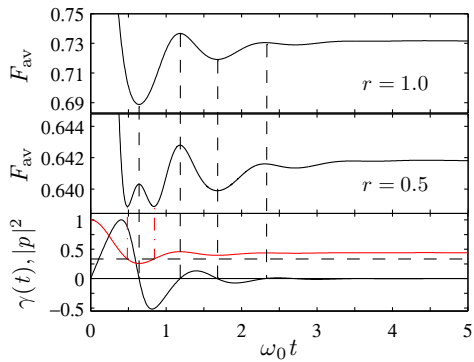}}
% If not, use\vspace{5cm}
% Give the correct figure height in cm
\caption{(Color online) Comparison of the behaviors of $F_{\rm av}$
         with that of the normalized decay rate $\gamma(t)$ (black) and the
         population parameter $|p(t)|^2$ (red) in super-Ohmic reservoir with
         $\omega_c=\omega_0$ and $\eta=0.9$. The horizontal dashed line
         represents constant 1/3.} \label{fig:5}
% Give a unique label
\end{figure}

To further understand the relations between dynamical behaviors of
$F_{\rm av}$ and $\gamma(t)$, we reexamine Eq. \eqref{eq11}, from
which one can derive $\Gamma(t)=-\dot{q}(t)/q(t)$, with
$q(t)=|p(t)|^2$. Therefore $\gamma(t)>0$ when $|p(t)|$ decreases
with time, and $\gamma(t)<0$ when $|p(t)|$ increases with time. As
$F_{\rm av}$ is a monotonic increasing function of $|p(t)|^2$ in the
region of $|p|^2>|p|_c^2=(1-r)/(1+r)$, $F_{\rm av}$ is increased
whenever $\gamma(t)<0$, and decreased whenever $\gamma(t)>0$ during
this region. In the region of $|p|^2<|p|_c^2$, however, $F_{\rm av}$
is a monotonic decreasing function of $|p(t)|^2$, thus $F_{\rm av}$
is decreased whenever $\gamma(t)<0$, and increased whenever
$\gamma(t)>0$. But it should be note that for this case (i.e.,
$|p|^2<|p|_c^2$), $F_{\rm av}$ cannot be larger than its classical
limiting value 2/3.

In Fig. \ref{fig:5} we gave an exemplified plot of the time
dependence of $F_{\rm av}$ and $\gamma(t)$ certifying explicitly the
above general arguments. When $r=1$, as mentioned above, $F_{\rm
av}$ behaves as a monotonic increasing function of $|p(t)|^2$ in the
whole time region, thus it is increased when $\gamma(t)<0$ and
decreased when $\gamma(t)>0$. For $r=0.5$ we have
$|p|^2<|p|_c^2=1/3$ during the time region
$\omega_{0}t\in(0.49,0.84)$, and from the solid red line shown in
Fig. \ref{fig:5}, one can see obviously that $F_{\rm av}$ is
decreased if $\gamma(t)<0$ and increased if $\gamma(t)>0$. As the
negativity of the decay rate is a signature of strong backaction of
the non-Markovian memory effect of the reservoir, the decrement of
$F_{\rm av}$ with $\gamma(t)<0$ indicates that the back flow of
information from the environment to the channel does not always
improve quality of teleportation.

\section{Summary}
In summary, we have studied quantum teleportation of a general
one-qubit state in the bosonic structured reservoirs with spectral
densities of the super-Ohmic, Ohmic and sub-Ohmic types. Through
comparing dynamical behaviors of the average teleportation fidelity
$F_{\rm av}$ with different system-reservoir parameters, we show
that the structures of the reservoirs play a decisive role on
quality of teleportation, and $F_{\rm av}$ can be improved by the
technique of reservoir engineering, e.g., by tuning the reservoir
structures so that there are bound state formulated. We have also
compared behaviors of $F_{\rm av}$ with that of the decay rate
$\gamma(t)$, and revealed the relationships between the
negativity/positivity of $\gamma(t)$ and the enhancement/decrement
of $F_{\rm av}$.

As it is now possible to simulate and control the non-Markovian
effect \cite{Almeida,Xujs,Liubh}, we hope our results may be
certified with state-of-the-art experiments, and at the same time
shed some new light for exploring the relationship between
non-Markovianity of the reservoir and the efficiency for performing
certain quantum protocols.

\begin{center}
\textbf{ACKNOWLEDGMENTS}
\end{center}

This work was supported by NSFC (11205121, 11174165, 11275099), and
the Scientific Research Program of Education Department of Shaanxi
Provincial Government (12JK0986).

\newcommand{\PRL}{Phys. Rev. Lett. }
\newcommand{\PRA}{Phys. Rev. A }
\newcommand{\RMP}{Rev. Mod. Phys. }
\newcommand{\JPA}{J. Phys. A }
\newcommand{\JPB}{J. Phys. B }
\newcommand{\PLA}{Phys. Lett. A }
%

% BibTeX users please use
% \bibliographystyle{}
% \bibliography{}

\end{document}